\documentclass[twoside]{article}
\usepackage{Proc_NTSE_16}
\usepackage{epstopdf}
\usepackage{dsfont}

\begin{document}
\thispagestyle{plain}
\publref{XZhaoWDu}

\begin{center}
{\Large \bf \strut
%Insert the title of your contribution here
Deuteron Coulomb Excitation in Peripheral Collisions with a Heavy Ion
\strut}\\
\vspace{10mm}
{\large \bf 
%Insert the authors here. Use upper indexes a, b, c, etc., to bind authors with their addresses
% as shown below.
W. Du$^{a}$, P. Yin$^{b}$, G. Chen$^{a}$, X. Zhao$^b$ and J. P. Vary{$^a$}}
\end{center}

\noindent{\small $^a$\it Physics Department, Iowa State University, Ames, Iowa, U.S.A., 50010} \\
{\small $^b$\it Institute of Modern Physics, Chinese Academy of Sciences, Lanzhou, China, 730000} \\

%The next command defines running titles:
\markboth{
%Put here the list of authors that will be displayed in running titles:
W. Du, P. Yin, G. Chen, X. Zhao and J. P. Vary}
{%Put here the short title of your contribution that will be displayed in running titles:
Deuteron Coulomb Excitation by Heavy Ion} 

\begin{abstract}
We develop and test an {\it ab initio} time-dependent Basis Function (tBF) method to solve non-perturbative and time-dependent problems in quantum mechanics. For our test problem, we apply this method to the Coulomb excitation of the deuteron by an impinging heavy ion. In the tBF method applied to deuterium, we employ wave functions for its bound and excited states to calculate its transition probabilities and the r.m.s. radius during the scattering process. For comparison, corresponding results based on first-order perturbation theory are also provided. For the Coulomb excitation process in a weak and time-varying Coulomb field, where higher-order effects are negligible, we obtain good agreement of the results based on these two methods. The tBF method is then applied to the Coulomb excitation process with stronger external field. The higher-order effects, such as those appearing in the  reorientation of the polarization of the deuteron system, are analyzed.
\\[\baselineskip] 
\end{abstract}

\section{Introduction}
The importance of Coulomb excitation and its application in nuclear physics are well known (see, e.g., Ref. \cite{KAlder56} and references therein). A target nucleus transitions to excited states when scattered by the electromagnetic (E$\&$M) field produced by a projectile heavy ion. First order perturbation theory works well when the field is weak. When the field is strong (from, e.g., a highly charged ion), higher order effects, such as reorientation, become important. For a precise description, the numerical solution obtained from direct treatment of the time-dependent Schr$\ddot{o}$dinger equation is necessary \cite{JEisenberg88}. In this work, we present a non-perturbative method to solve the time-dependent Schr$\ddot{o}$dinger equation, called the time-dependent basis function (tBF) method. It is closely related to our previous work on time-dependent Basis Light-Front Quantization (tBLFQ) \cite{XZhao13, Chen:2017uuq}. It enables tracking the evolution of quantum states as a function of time. The dynamics of the quantum system is revealed at the amplitude level. The tBF method will be especially useful when the interactions are strong, in which cases the perturbative calculations are not reliable.

In this paper, the deuteron Coulomb excitation problem is studied as a first test. The heavy ion impinges along a fixed impact parameter so the center of mass of the deuteron is held fixed during the collision process. The impact parameter is set to be sufficiently large such that the strong nuclear force does not affect the scattering. The projectile ion generates both time-varying electric (Coulomb) field and magnetic field. While both the neutron and the proton interact with the magnetic field, only the proton gets repelled by the Coulomb field.
%The proton is repelled by the time-varying Coulomb field, while both the neutron and the proton interact with the time-varying magnetic field, both generated by the projectile ion. 
After the scattering, the polarization of the deuteron system has been reoriented. For our initial test application, we will consider only the Coulomb excitation effects modeled through the electric dipole transition operator. 

\section{Theory and Properties of the Test Application}

For the purposes of introducing our tBF approach, we will take the specific example of a peripheral heavy-ion collision with a deuteron and consider only its Coulomb excitation.  The target can easily be generalized to other systems such as $^6$Li or $^{12}$C. Each of the simplifications will be lifted in future efforts as our main purpose here is to introduce the method and provide initial tests.  A main feature of our approach is that we employ {\it ab initio} solutions for the ground and excited states of the target system based on a realistic inter-nucleon interaction. Of course, phenomenological target wave functions may also be employed and may be necessary for heavier targets.

\subsection{Target Properties}
Within our application of the tBF method, the state wave functions of the target, the neutron-proton (np) system, are solved with JISP16 NN-interaction \cite{AShi05, AShi07, AShi04}. A multipole expansion is conducted for the Coulomb field \cite{ABohr} produced by the heavy ion and only the E1 multipole component is taken into account. The interaction between the field and the np system can be expressed in terms of the nuclear matrix elements that determine the radiative transitions. The time-dependent state wave functions of the np system are solved numerically and then applied to calculate the transition probabilities between states as well as the r.m.s. radius of the np system. We demonstrate how the tBF method provides a complete dynamic picture of the evolution of the np system, where all the higher-order effects are taken into account.

\subsection{Peripheral Scattering Setup}
\begin{figure}[ht]
\centering
\includegraphics[width=13cm]{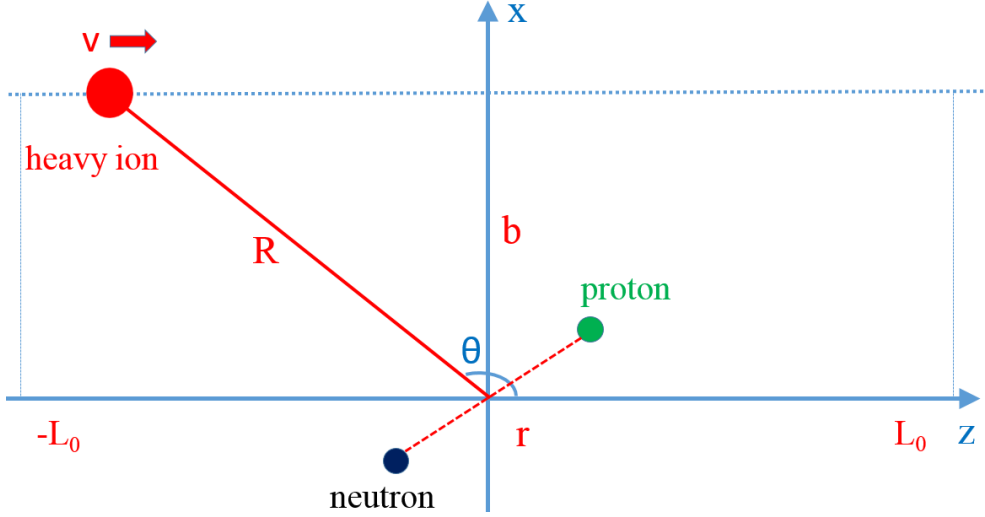}
\caption{The schematic view of the scattering between the heavy ion and the deuteron target. The scattering plane is set to be in the $x$-$z$ plane. The heavy ion moves straight along the positive $\hat{z}$-axis with a constant speed. The center of mass of the target is fixed at the origin. No recoil of the target is considered when the heavy ion skims over. $b$ is the impact parameter, $v$ is the speed of the heavy ion projectile (along the $\hat{z}$-axis), $L_0$ is the horizontal distance to cut off the Coulomb interaction between the np system and the heavy ion, $\theta$ is the polar angle of the position of the heavy ion, $R$ is the distance from the center of heavy ion to the origin, $r$ is the separation between the proton and the neutron.
}
\label{fig:SetUP1}
\end{figure}

The scattering plane is set to be the $x$-$z$ plane, as shown in Fig. \ref{fig:SetUP1}. The heavy ion, with charge $Ze$, moves with a constant speed (taken as 0.1c for our numerical results) along the $\hat{z}$-axis. The impact parameter is sufficiently large (taken as 7.5 fm for our numerical application) such that the strong nuclear force does not play a role. The nucleon mass is taken to be $938.92$ MeV and the unit charge of the deuteron target is carried by the proton. No meson exchange current is evaluated. As an approximation, we consider the case where the impact parameter is constant throughout the collision process so that the center of mass of the target is always fixed at the origin. In this work, we consider only the E1 multipole contribution for the Coulomb excitation, though other multipole components, e.g., E0, M1, E2 etc., contribute to the full problem. A cut-off distance is introduced for the Coulomb interaction, beyond which no significant transition takes place.

\subsection{Hamiltonian}
In the relative coordinates of the np system, the full Hamiltonian $H_{full}$ (for the target interacting with external E$ \& $M field generated by the moving heavy ion) consists of two parts 
\begin{eqnarray}
H_{full} &=& H_0 + V_{int} \label{eq:FullH} \ ,
\end{eqnarray}
where
\begin{eqnarray}
H_0 &=& H_{KE} + V_{NN} + H_{ext} \label{eq:H0}
\end{eqnarray}
is the time-independent Hamiltonian for the np system. The $H_{KE}$ is the intrinsic kinetic energy of the np system. The $V_{NN}$ describes the realistic nucleon-nucleon (NN) interaction. In this work, it is taken to be the JISP16 NN interaction \cite{AShi05, AShi07, AShi04}. Since the np system is weakly bound, a small external harmonic oscillator (HO) potential $H_{ext}$ with strength $ \hbar \omega = 5$ MeV is introduced to regulate the continuum states and produce a discretized representation of the continuum.  In future works, we will employ a basis space regulator that does not affect the ground state while discretizing the continuum without an external field. 

All the eigenstates of $H_{0}$ are obtained by diagonalization in a sufficiently large model space
\begin{eqnarray}
H_{0} |\beta_j \rangle &=& E_{j} \ |\beta_j \rangle \label{eq:EFunc} .
\end{eqnarray}
These energy eigenstates form the complete basis set $\{|\beta_j \rangle \}$ of the np system in our test problem. The time-dependent background field $V_{int}$ induces transitions between eigenstates of the np system in the trap. 

\subsection{Background Field}
$V_{int}$ is the time-dependent part of the full Hamiltonian. It is evaluated locally through the coupling between the four potential $A^{\mu}=(\varphi,\ \vec{A})$ from the moving charge and the four current of the np system $J^{\mu}=(\rho,\ \vec{j})$ 
\begin{eqnarray}
V_{int}(t) &=& \int A_{\mu}J^{\mu} \ d \vec{r} \ = \ \int \rho (\vec{r},t) \varphi(\vec{r},t) \ d \vec{r} \ -\ \int \vec{j}(\vec{r},t) \cdot \vec{A}(\vec{r},t) \ d \vec{r} \ \label{eq:CouplingEnergy},
\end{eqnarray}
where the relative coordinates relate to the single-particle coordinates of the np system as $\vec{r}=\vec{r}_p - \vec{r}_n$. The second term in Eq. \eqref{eq:CouplingEnergy} is neglected since the heavy ion is moving with low velocity in our initial application. Only the interaction between the Coulomb field of the incident heavy ion and the charge of the np system is evaluated. Moreover, the multipole expansion of the Coulomb field \cite{ABohr} is performed and only the E1 term is kept.

\subsection{Interaction Picture}
The basis set is formed by the eigenstates of the free Hamiltonian $H_0$. The transitions between these eigenstates are described in the interaction picture, in which the equation of motion (EOM) of the np system is 
\begin{eqnarray}
i \frac{\partial}{\partial t}|\psi; t \rangle _I &=&  e^{i {H_{0}t}}\ V_{int}(t)\ e^{-i {H_{0}}t}\ |\psi; t \rangle _I \ \equiv \  V_I(t)\ |\psi; t \rangle _I \ ,
\end{eqnarray}
where $V_I(t)$ is the interaction part of Hamiltonian in the interaction picture. The subscript ``I'' is adopted to distinguish the quantities in the interaction picture from those in the Schr$\ddot{o}$dinger picture. The above EOM can be solved by integration
\begin{eqnarray}
|\psi; \ t \rangle _I &=& \hat{T} \Bigg\{  \exp \Bigg[-i\int_{t_0}^t \ V_I(t')\ dt' \Bigg]\Bigg\} |\psi;\ t_0 \rangle _I \label{eq:EOMsoln} \ ,
\end{eqnarray} 
where $\hat{T}$ is the time ordering operator towards the future. Instead of functional expansion in perturbation theory, the above equation is evaluated numerically in the tBF approach. For this purpose, we divide the time interval $[-T,\ T]$ into n segments, each segment with step length $\delta t=\frac{2T}{n}$. The integration in the exponent is then replaced as
\begin{align}
\hat{T}\Bigg\{  &  \exp\Bigg[-i\int_{-T}^T \ V_I(t)\ dt \Bigg]\Bigg\} \nonumber \\ 
&\xrightarrow{\sum \delta t} \Bigg[ 1-i\ V_I(t_n)\delta t \Bigg]\ \Bigg[ 1-i\ V_I(t_{n-1})\delta t \Bigg]\ \cdots \ \Bigg[ 1-i\ V_I(t_1)\delta t \Bigg] \label{eq:Evolv} \ .
\end{align}
By multiple insertions of the projection operator defined from this complete basis set in Eq. \eqref{eq:EFunc}
\begin{equation}
\mathds{1} = \sum_{j} |\beta_j \rangle \langle \beta_j | \ ,
\end{equation}
the right hand side of Eq. \eqref{eq:Evolv} reduces to matrix multiplications. The final  state after evolution $|\psi; \ t \rangle _ I$ in Eq. \eqref{eq:EOMsoln} is therefore obtained.

\subsection{Interaction Matrix}
Taking only the E1 multipole component of the Coulomb field, the transition matrix element \cite{KAlder56, JEisenberg88} is
\begin{eqnarray}
\langle \beta _j | V_I(t) | \beta _k \rangle &=& \frac{4 \pi}{3} Ze^2 e^{i({E_{j} - E_{k}})t} \sum_{\mu} \ \frac{Y^{\ast}_{1 \mu}(\Omega_{\hat{R}})}{|R(t) |^{2}} \langle \beta _j | \frac{r}{2}Y_{1 \mu}(\Omega_{\hat{r}}) | \beta _k \rangle \label{eq:IntMatElem} \ ,
\end{eqnarray}
where $E_j$ and $E_k$ are respective eigenenergies of eigenvectors  $| \beta _j \rangle$ and $| \beta _k \rangle$ of the Hamiltonian $H_0$. $Y_{\lambda \mu}(\Omega)$ ($\lambda = 1$ for the E1 multipole contribution) denotes the spherical harmonics following the Condon-Shortley convention \cite{JSuhonen}. $Ze$ is the charge of the heavy ion while the np system carries charge $e$ in total.  In the relative coordinates of the np system, $r$ is the distance from the proton to the neutron, while $\vec{R}(t)$ is the position of the heavy ion projectile. The kernel in Eq. (9) is the matrix element for the E1 transition between eigenstates of the np system.  

To compute the interaction matrix, the three dimensional harmonic oscillator (3DHO) representation is adopted. We adopt a model space truncation parameter $2n+l \le N_\text{max}=60$ to define our approximation to the full basis space which leads to our definition of  the complete 3DHO basis set $\{ |n l s JM \rangle \}$ for each eigenvector in the basis set $\{| \beta_j \rangle \}$ of the np system (specified by good quantum numbers $s$, $J$ and $M$)   
\begin{eqnarray}
\mathds{1} &=& \sum_{nl} |nlsJM  \rangle \langle nlsJM| \ .
\end{eqnarray}
Here, for each 3DHO basis wave function, $n$ is the radial quantum number which denotes the number of nodes of the radial part of the wave function, $l$ is the quantum number of orbital angular momentum $\vec{l}$, $s$ is the quantum number for spin $\vec{s}$, $\vec{l}$ and $\vec{s}$ couple to the total angular momentum $\vec{J}$, which is a conserved quantity of the Hamiltonian $H_0$. $M$ is the magnetic quantum number of $\vec{J}$ along the quantization axis ($\hat{z}$-axis in this case).
The E1 matrix element in Eq. \eqref{eq:IntMatElem} becomes
\begin{align}
& \langle \beta _j | \frac{r}{2}Y_{1 \mu}(\hat{r}) | \beta _k \rangle  \nonumber \\
=&  \sum_{nl} \sum_{n'l'} \langle \xi_j J_j M_j |nlsJ_jM_j \rangle \langle nlsJ_jM_j | \frac{r}{2}Y_{1 \mu}(\hat{r})| n'l's'J_k M_k \rangle \langle n'l's'J_k M_k | \xi_k J_k M_k \rangle \ , \label{eq:E1kernel}
\end{align}
where 
\begin{eqnarray}
| \beta _j \rangle  &=& \sum_{nl} \langle nlsJ_j M_j| \xi_j J_j M_j \rangle | nlsJ_j M_j \rangle \ \equiv \ \sum_{nl} a_{j;nl} | nlsJ_j M_j \rangle , \label{eq:21} \\
| \beta _k \rangle &=& \sum_{n'l'} \langle n'l's'J_k M_k| \xi_k J_k M_k \rangle | n'l's'J_k M_k \rangle \ \equiv \ \sum_{n'l'}a_{k;n'l'} | n'l's'J_k M_k \rangle . \label{eq:20}
\end{eqnarray}
and $\xi_j$ and $\xi_k$ denote any additional quantum numbers necessary to describe $| \beta _j \rangle$ and $| \beta _k \rangle$, respectively. In this work, the amplitudes $\{a_{j;nl} \}$ and $\{a_{k;n'l'} \}$ are solved by diagonalization of $H_0$ in the 3DHO representation. The middle kernel in Eq. \eqref{eq:E1kernel} is the E1 matrix element $({r}/{2})Y_{1 \mu} (\hat{r})$ in the 3DHO representation. It  can be solved by converting into the coordinate representation 
\begin{align}
& \langle nlsJ_j M_j | \frac{r}{2}Y_{1 \mu} (\hat{r})| n'l's'J_k M_k \rangle  \nonumber \\
=& \sum_{m_l m_s} \sum_{m_l' m_s'} \delta_{ss'} \delta_{m_s m_s'} (lm_lsm_s|J_j M_j)(l'm_l's'm_s'|J_k M_k)  \int R_{nl}^{\ast}(r) \frac{r}{2} R_{n'l'}(r)r^2 dr \nonumber \\ 
& \times (-1)^{m_l} \sqrt{\frac{3(2l+1)(2l'+1)}{4\pi}}
\begin{pmatrix}
l & 1 & l' \\ 
-m_l & \mu & m_l'
\end{pmatrix} 
\begin{pmatrix}
l & 1 & l' \\ 
0 & 0 & 0
\end{pmatrix} , \label{eq:kernel}
\end{align}
where $R_{nl}(r)$ is the radial part of 3DHO wave function in the coordinate representation
\begin{eqnarray}
R_{nl}(r) &=& \sqrt{\frac{2n!}{r_0 \Gamma(n+l+\frac{3}{2})}} \ \Big( \frac{r}{r_0} \Big)^{l+1} \ \exp\Big[ -\frac{r^2}{2 r_0^2} \Big] \ L_n^{l+\frac{1}{2}}\Big( \frac{r^2}{ r_0^2} \Big) \ ,
\end{eqnarray}
with $L_{n}^{\alpha}({r^2}/{ r_0^2})$ the associated Laguerre polynomial. $r_0=\sqrt{1/{m \omega}}$ is the oscillator length with $ m $ the reduced mass of the np system and $ \omega$ the HO frequency taken to be the same as the frequency of the trap. This definition ensures that $R_{nl}(r)$ starts positive at the origin. The radial integral in Eq. \eqref{eq:kernel} is evaluated to be
\begin{align}
& \int R_{nl}^{\ast}(r) \frac{r}{2} R_{n'l'}(r)r^2 dr \nonumber \\
=& \frac{r_0}{2} \left\{
\begin{array}{c}
\sqrt{n+l+\frac{3}{2}}\ \delta_{nn'} - \sqrt{n} \delta_{n,n'+1} \ \ \ \ \  \text{for} \ l<l' \ \text{and}\ n \geq n' \\ 
\sqrt{n'+l'+\frac{3}{2}}\ \delta_{nn'} - \sqrt{n'} \delta_{n',n+1} \  \ \text{for} \ l'<l \ \text{and}\  n' \geq n \\ 
0 \ \ \ \ \ \ \ \ \ \ \ \ \ \ \ \ \ \ \ \ \ \ \ \   \ \ \text{for}\  l=l' \ 
\end{array} \right. \ .
\end{align}
$(lm_lsm_s|J_jM_j)$ in Eq. \eqref{eq:kernel} is the CG-coefficient and $$\begin{pmatrix}
l & 1 & l' \\ 
-m_l & \mu & m_l'
\end{pmatrix}
$$ is the $3j$-symbol following the Condon-Shortley convention \cite{JSuhonen}. The angular part determines the selection rule of the E1 transition.

\subsection{Evolution of States}
The method described in Eq. \eqref{eq:Evolv} is known as the Euler scheme. This approach is not stable because it is not symmetric in time. The norm of the state vector $|\psi; t>_I $ may not be conserved under time evolution  \cite{AAskar78}, which violates the conservation of probability. We therefore adopt the MSD2 \cite{TIitaka94} scheme
\begin{eqnarray}
|\psi, t+\delta t \rangle _I &\approx & |\psi, t-\delta t\rangle _I -{2i} \ V_I(t) \ \delta t \ |\psi, t\rangle _I \label{MSD2} \ .
\end{eqnarray}

For comparison with results from the MSD2 scheme, we also present the evolution calculated from the first-order perturbation theory. From Eq. \eqref{eq:Evolv}, the evolution of state vector is evaluated to the leading order in the interaction $V_I$, 
\begin{eqnarray}
|\psi; \ t\rangle _I  &\rightarrow&  \Bigg[1 -i\ \delta t \Big(V_I(t_n)+V_I(t_{n-1})+ \cdots +V_I(t_1) \Big) \Bigg] |\psi;\ 0\rangle _I \ .
\end{eqnarray}

\subsection{Observables}
During the evolution, the wave function of the np system at a certain moment in terms of the basis set $\{|\beta_j \rangle \}$ is
\begin{eqnarray}
|\psi; \ t \rangle  &=& \sum_{j} A_{j}(t) | \beta _{j} \rangle  \ \label{eq:wfunction},
\end{eqnarray}
where $  | \beta _{j} \rangle  $ denotes the tBF basis solved from Eq. \eqref{eq:EFunc}. The amplitudes $A_j(t)$ are tracked for each basis state during evolution. Applying the time-dependent np wave function Eq. \eqref{eq:wfunction}, we can track the r.m.s. radius of the np system as
\begin{eqnarray}
 \langle r^2 \rangle ^{\frac{1}{2}} &=& \frac{1}{2} \sqrt{ \langle \psi; \ t |\ {r}^2 \ | \psi; \ t \rangle } \ .
\end{eqnarray}

\section{Results and Discussion}

\begin{figure}[ht]
\centering
\includegraphics[width=8cm]{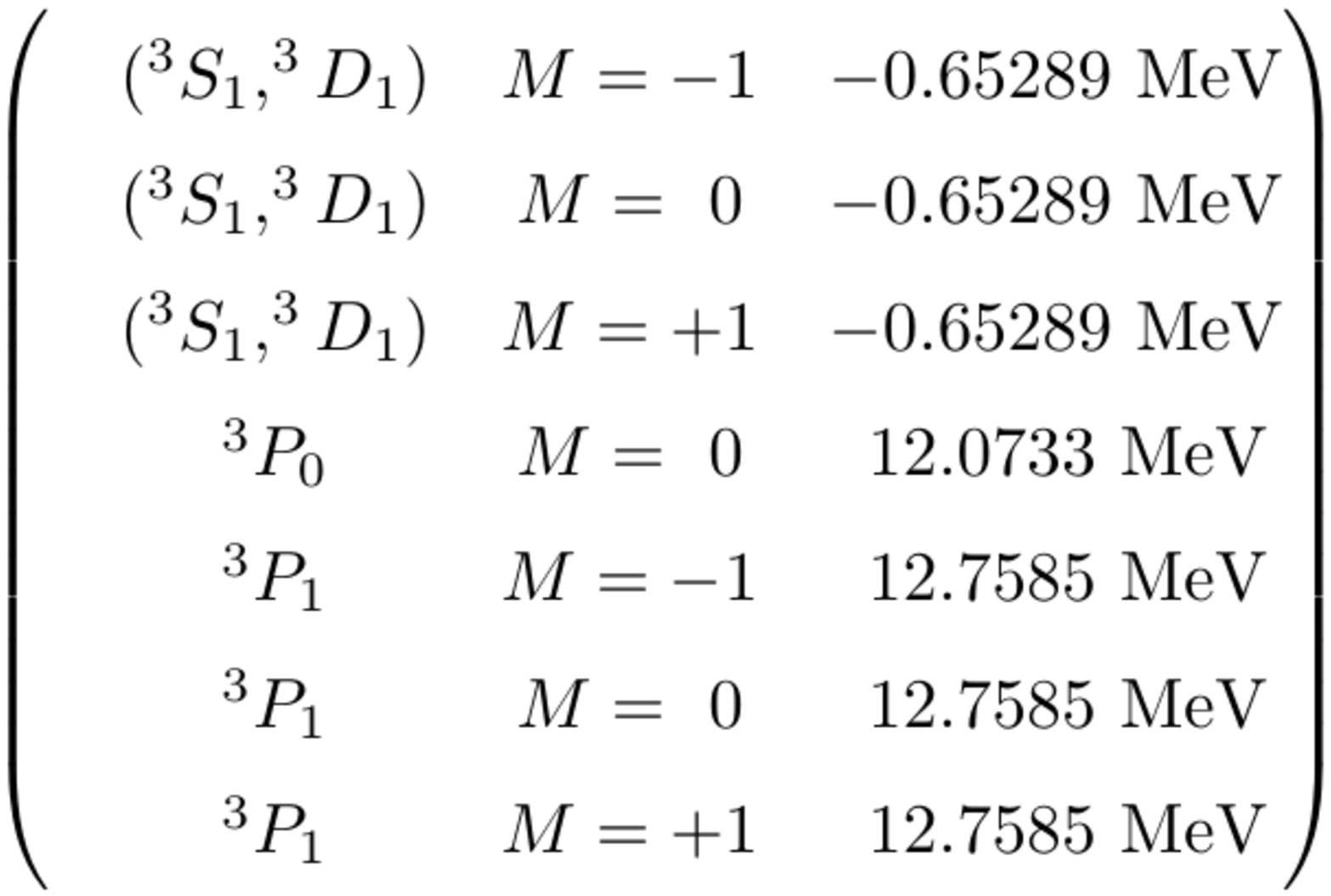}
\caption{Basis set of the np system for evolution.}
\label{fig:basism}
\end{figure}

For our test problem, we consider 3 interaction channels for the np system. They are $(\ ^3S_1, \ ^3D_1)$, $ \ ^3P_0$ and $\ ^3P_1$. For each channel, the lowest states (degenerate in magnetic quantum numbers $M$) of $H_0$ are considered. The eigenenergy of each state is also shown in Fig. \ref{fig:basism}.

The heavy ion projectile is chosen to be fully stripped (all electrons removed) $^{124}\text{Sn}$, with $Z=50$. It moves straight along the positive $\hat{z}$-axis with a constant speed $0.1$c. The center of mass of the np system is fixed at the origin. The impact parameter is fixed to be $7.5$ fm. The exposure duration is $10 \ \text{MeV}^{-1}$ ($6.582 \times 10^{-21}$ sec), during which time the projectile travels approximately 200 fm (100 fm before distance of closest approach and 100 fm after that). The initial state is prepared to be polarized along the negative $\hat{z}$-axis. That is, the initial state is selected to be the $(\ ^3S_1, \ ^3D_1), M=-1$ state. 

Applying either the tBF method or the first-order perturbation theory, we will calculate the wave functions of the np system at selected moments during evolution. The wave functions are then applied to calculate the transition probabilities as well as the r.m.s. radius of the np system as a function of time. The dependence of these properties on the strength of external Coulomb field is also investigated by altering the fine-structure constant.  We note that our discussion of properties at intermediate times is allowed in quantum mechanics but only the results at asymptotic times correspond to experimental observables.

\begin{figure}[ht]
\centering
\includegraphics[width=13cm]{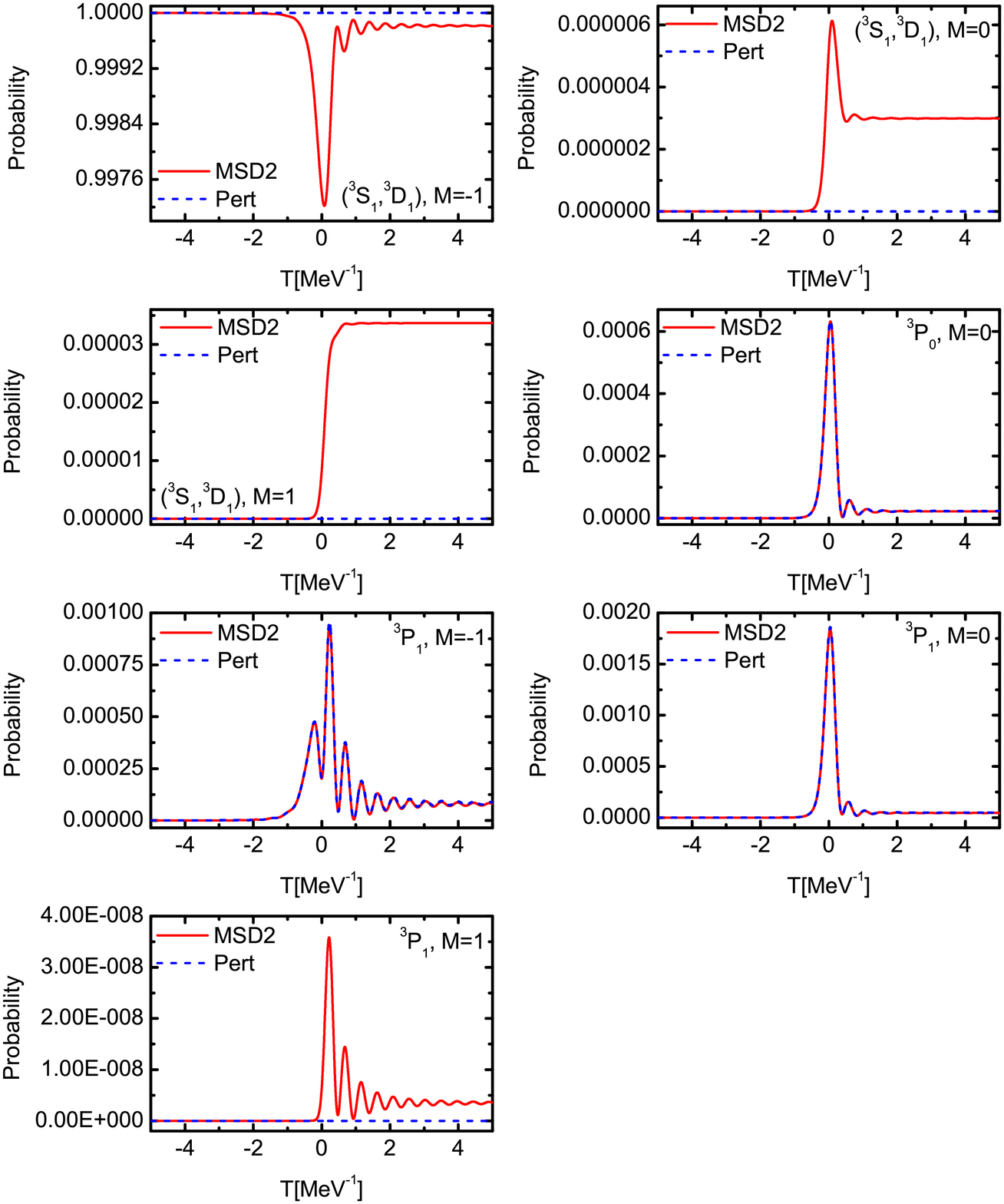}
\caption{Time evolution of the np system (characterized by 7 basis states) in the weak and time-varying Coulomb field. The initial deuteron system is polarized along the negative $ \hat{z} $-axis. The duration is $10 \ \text{MeV}^{-1}$ and the fine-structure constant is ${1}/{137.04}$. The charge of the heavy ion projectile is $Z=50$ and it moves along the positive $\hat{z}$-axis with a constant speed $0.1c$. The impact parameter is $7.5$ fm. The initial state is prepared to be the $ (^3S_1, \ ^3D_1), M=-1$ ground state (the lowest basis state in our eigenbasis). For each basis state, the red solid curve represents the probability calculated by the MSD2 scheme during evolution, while the blue dashed curve is the result from first-order perturbation theory.}
\label{fig:2T10weakNegZ}
\end{figure}

\subsection{Weak Field Results}
In Fig. \ref{fig:2T10weakNegZ}, we plot the time evolution of the probabilities of the np system when it is exposed to a weak, time-varying Coulomb field. The coupling constant is set to be $\alpha={1}/{137.04}$. 

Intense probability fluctuations are found for some states at the middle of the scattering process, when the heavy ion is close to the np system. Such fluctuations are transient and they are signs of the virtual quantum processes. The probabilities converge to stable values when the Coulomb field fades away. The distance for the Coulomb field to be effective is related to the external interaction energy $V_{int}$ and internal energy gaps of the np system.  

For levels that obey the E1 selection rule, good agreement for the transition probabilities is obtained between calculations from the tBF method (red solid lines) and calculations from first-order perturbation theory (blue dashed lines). This shows that the tBF method is consistent with first-order perturbation theory when the interaction field is weak. For states that violate the E1 selection rule, however, differences are found between the results from the tBF method and those from first-order perturbation theory. For example, the forbidden states, $ (^3S_1, \ ^3D_1), M=0$ state, $ (^3S_1, \ ^3D_1),M=1$ state and $ \ ^3P_1, M=1$ state are excited at the end of evolution, though with relatively small probabilities. These ``forbidden transitions" result from the higher-order processes which are excluded from first-order perturbation theory. 

\begin{figure}[ht]
\centering
\includegraphics[width=13cm]{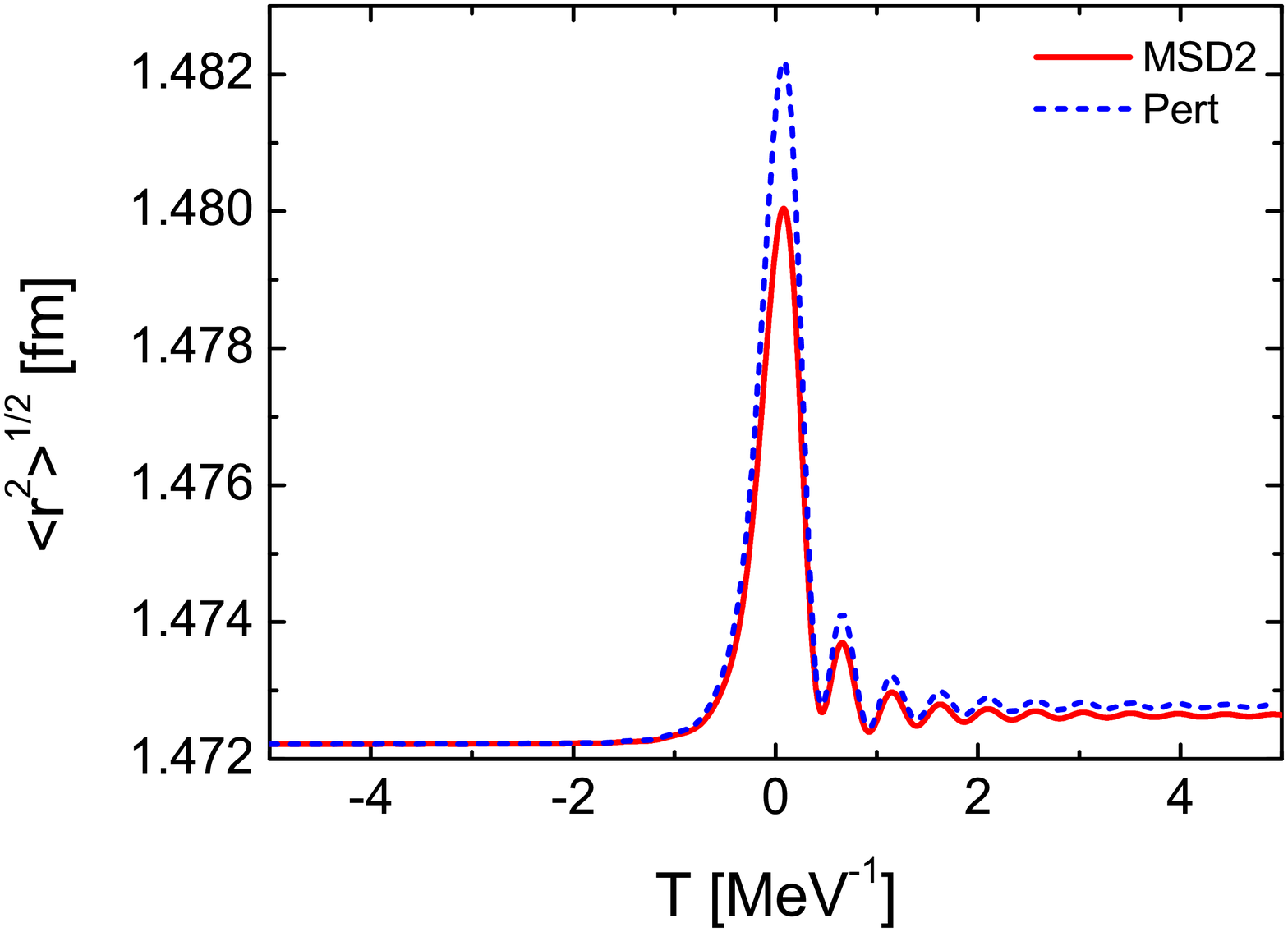}
\caption{The r.m.s. radius of the np system during evolution. The simulation conditions for obtaining the intermediate np wave functions are the same as those described in Fig. \ref{fig:2T10weakNegZ}. The red solid curve represents the r.m.s. radius calculated from the wave function obtained via the tBF method, while the blue dashed curve is the r.m.s. radius based on the wave function from first-order perturbation theory.}
\label{fig:RMS2T10weakNegZ}
\end{figure}

In Fig. \ref{fig:RMS2T10weakNegZ}, we present the r.m.s. radius for the np system during evolution. Due to the HO potential we choose to constrain the np system, its r.m.s. radius, 1.472 fm, is approximately 25$\%$ smaller than the r.m.s. radius, 1.975(3) fm, of the physical deuteron \cite{Huber:1998zz, {Martorell:1995zz}}. The r.m.s. radius expands when the np system gets excited to high-lying levels. The tiny difference in the r.m.s. radii given by the two approaches is due to the ``forbidden transitions" to high-lying levels, which are the higher-order effects included by the tBF method. In general, the r.m.s. radius given by the tBF method (red solid line) agrees with that based on the perturbation theory (blue dashed line). At the end of evolution, both methods predict the net expansion of the order of $10^{-4} \ \text{fm}$.

\subsection{The Strong Field Results}
We test the tBF method for the evolution of the np system in the presence of a stronger and time-varying Coulomb field, where the fine structure constant is set to $\alpha=1/13.7$. We perform this calculation in order to enhance the visibility of the non-perturbative quantum effects that may become more evident with closer encounters and/or with relativistic heavy ions.

\begin{figure}[ht]
\centering
\includegraphics[width=13cm]{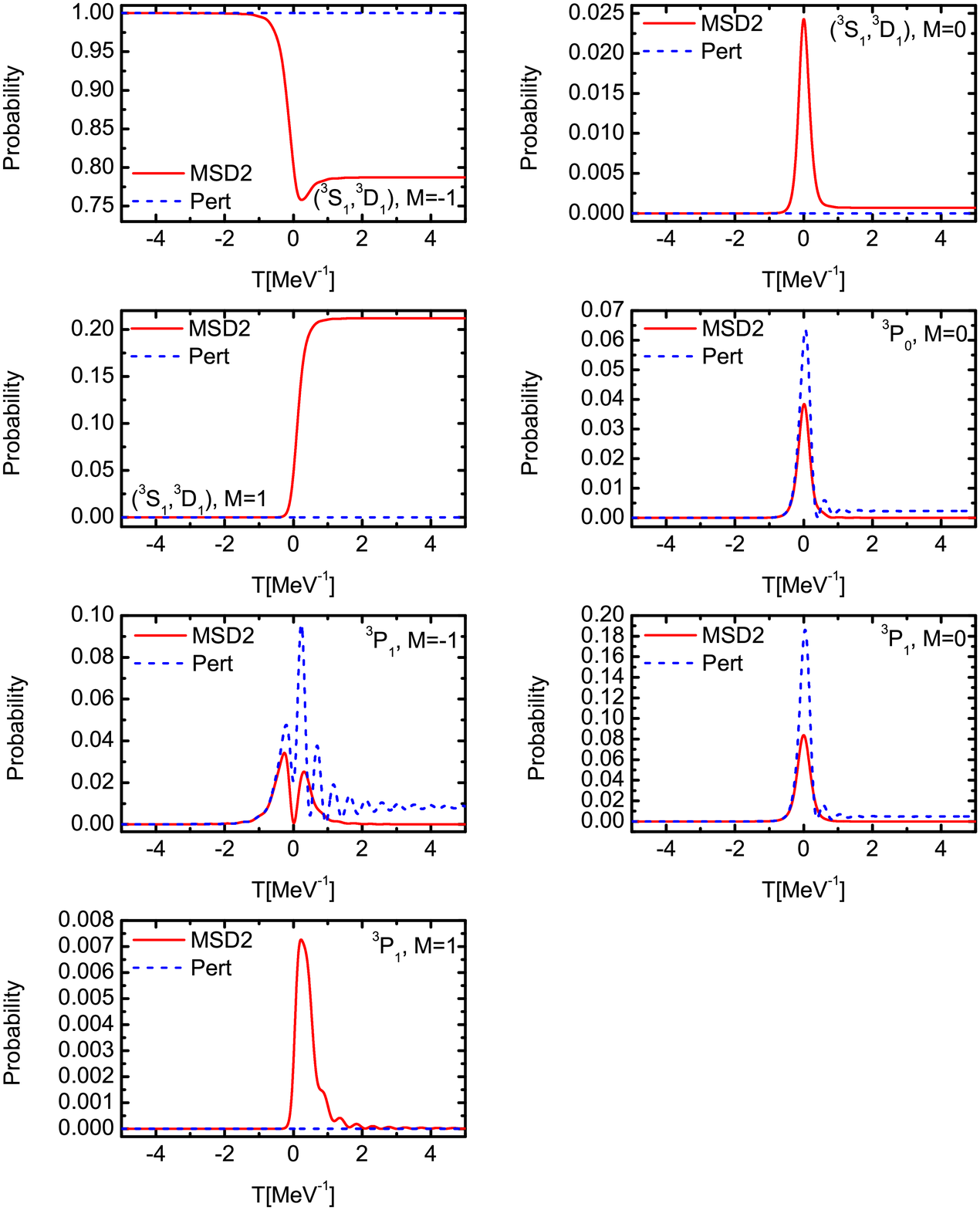}
\caption{The same as in Fig. \ref{fig:2T10weakNegZ}. However, the fine-structure constant is set to 1/13.704.}
\label{fig:2T10StrongNegZ}
\end{figure}

For the np system, the non-perturbative tBF calculation for the transition probabilities during evolution (red solid lines) are shown in Fig. \ref{fig:2T10StrongNegZ}. The transition probabilities based on first-order perturbation theory are again provided (blue dashed lines) for comparison. However, it is easily anticipated, and observed, that first-order perturbation theory is not sufficient in this case.

The evident difference on transition probabilities is found between the calculation based on the tBF method and that on first-order perturbation theory. This difference shows that the higher-order effects are crucial for the precise calculation of the Coulomb excitation process in a strong field. Such higher order effects result in a significant reorientation of the polarization of the np system. This is observed from the large transition probabilities to levels that are forbidden by the first order E1 selection rule, e.g., to the $ (^3S_1, \ ^3D_1), M=1$ state.  

\begin{figure}[ht]
\centering
\includegraphics[width=13cm]{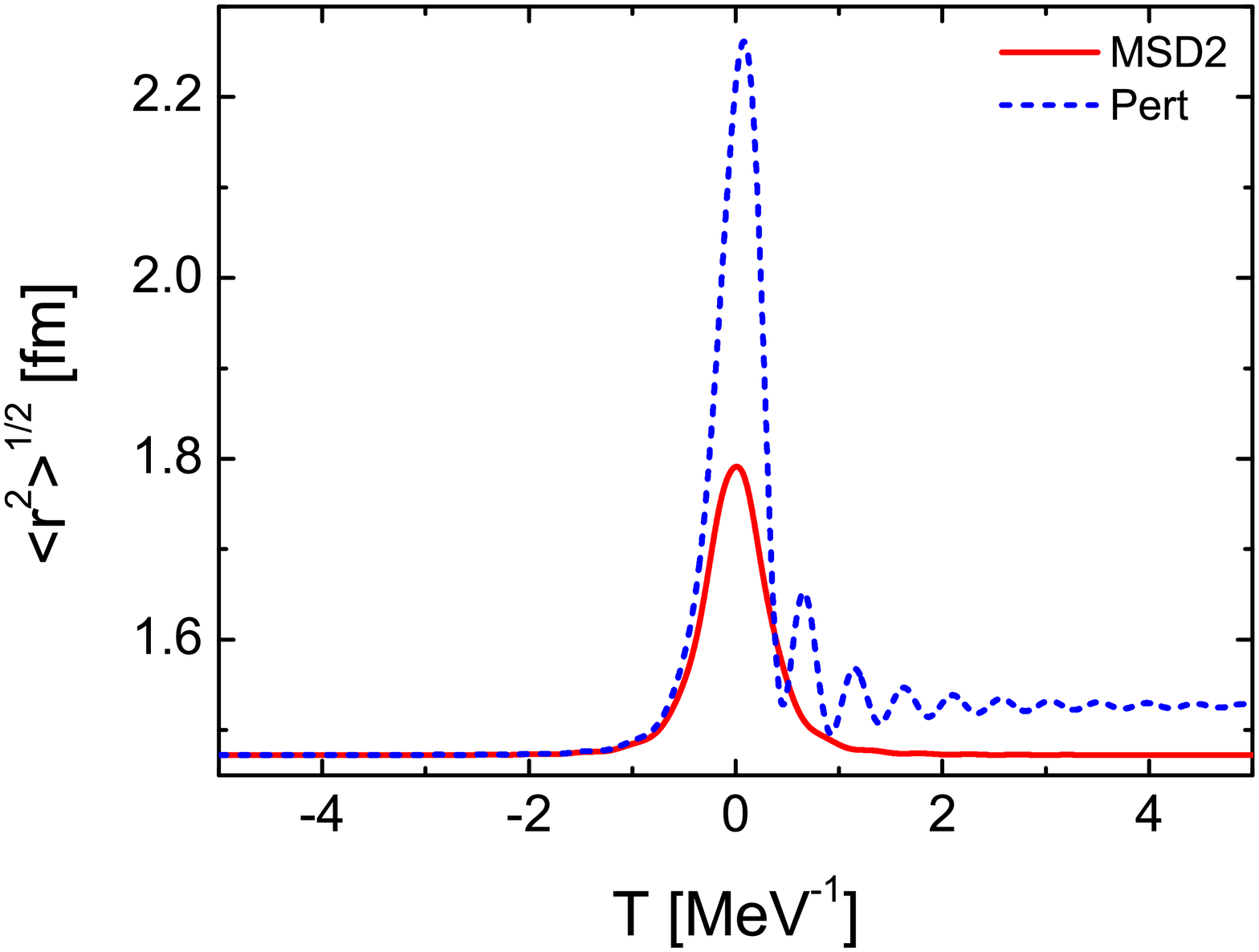}
\caption{The same as in Fig. \ref{fig:RMS2T10weakNegZ}. However, the fine-structure constant is set to 1/13.704.}
\label{fig:RMS2T10SrongNegZ}
\end{figure}

We observe that the np system is not significantly excited at the end of evolution. Though re-distributed, almost all of the population are still resident in the $ (^3S_1, \ ^3D_1)$ levels. This results in a negligible r.m.s. expansion of the np system after the scattering process (Fig.  \ref{fig:RMS2T10SrongNegZ}), which is at the order of $10^{-4} \ \text{fm}$. Note that the r.m.s. radius is a simple characteristic of the full final state distribution.  Population of the excited states would be expected to lead to breakup or gamma emission back to the ground state but we do not incorporate those final state effects in our current calculations.
%The magnitude of net expansion agrees with the previous r.m.s. calculation when a weak Coulomb field is present, which suggests that the net r.m.s. expansion is insensitive to the field-strength. This could be due to the limited basis size of the deuteron system applied in the current test problem. 

\section{Conclusions and Outlook}

We have developed and applied a non-perturbative method, the time-dependent Basis Function (tBF) method, to study scattering problems in strong and time-dependent external fields. Since the tBF method enables calculations of intermediate state wave functions, it enables a detailed investigation on the dynamics of a system during evolution. As a test problem, we study the Coulomb excitation of the deuteron target when a heavy ion projectile impinges. The target deuteron is placed in a weak external harmonic oscillator potential trap and its center of mass is fixed. The energy eigenfunctions of the target (np) system are solved with the JISP16 NN-interaction, from which only seven of them are kept as basis states for this test problem.
For simplicity, only the E1 component of the Coulomb field is considered. Due to the Coulomb excitation, the np system gets excited during the scattering process. We calculate the np wave functions at selected moments during evolution based on the tBF method. The wave functions are then applied to calculate the transition probabilities as well as the r.m.s. radius of the np system. In comparison, we also provide corresponding calculations which are based on first-order perturbation theory.

We first study the Coulomb excitation problem in a weak, time-varying external field. For those states that obey the E1 selection rule, we obtain agreement for the transition probability between these two approaches. This confirms that the tBF method is consistent with first-order perturbation theory in the limit of the weak interaction field. However, for the other states that violate the E1 selection rule, deviations from zero are obtained in the tBF approach. These deviations signify the higher-order effects missing in first-order perturbation theory.

In scattering problems with a stronger interaction field, the higher-order effects are expected to be important. To show this, we investigate the same Coulomb excitation problem but with a strong, time-varying external field.
We achieve this by tuning the fine-structure constant to $1/13.7$,  while all the remaining parameters are kept the same as those in the previous simulation. In this case, it is found that the higher-order effects largely reorient the polarization of the deuteron system, while the r.m.s. radius changes minimally after the evolution. At later times, differences in the level distribution and r.m.s. radius of the np system are observed between predictions based on the tBF method and first-order perturbation theory. This justifies a full non-perturbative treatment in the presence of strong interaction field.  

In the next step, with the validity of the tBF method confirmed, we will apply this method to simulate the scattering process of the deuteron in the presence of both the electromagnetic interaction and the strong interaction due to an impinging heavy ion. In such cases, both the Coulomb interaction and the strong interaction can modify the polarization of the deuteron system \cite{LOu2015, VBar2010, HSeyf2010}. Specific attention will be paid to the time-evolution of the charge and momentum distribution of the np system. These studies will be very important for understanding the dynamics of the deuteron breakup reaction \cite{CABert92, LFCanto97} and would serve as a precurser for investigating the internal structures of larger nuclei.

\section*{Acknowledgments}
We acknowledge valuable discussions and help from Y. Li, L. Ou and Z. Xiao. This work was supported in part by the U. S. Department of Energy (DOE) under grants No. DESC0008485 (SciDAC/NUCLEI) and DE-FG02-87ER40371. X. Zhao is supported by new faculty startup funding from the Institute of Modern Physics, Chinese Academy of Sciences.

\end{document}